\documentclass[proof]{pasj00}
\draft
\begin{document}

\title{Statistical Properties of Molecular Clumps \\in the Galactic Center 50 km s$^{-1}$ Molecular Cloud}
\SetRunningHead{Tsuboi et al.}{50 km s^{-1} Molecular Cloud}
\Received{2011 December 20}
\Accepted{2012 April 18}
\author{Masato T\sc{suboi}}
\affil{Institute of Space and Astronautical Science, Japan Aerospace Exploration Agency,\\
3-1-1 Yoshinodai, Chuo-ku, Sagamihara, Kanagawa 252-5210}
\affil{Department of Astronomy, the University of Tokyo, Bunkyo, Tokyo 113-0033}
\email{tsuboi@vsop.isas.jaxa.jp}
\and
\author{Atsushi M\sc{iyazaki}}
\affil{Korean VLBI Network, Korea Astronomy and Space Science Institute, \\
 P.0. Box 89, Yonsei University, 262 Seongsan-ro, Seodaemun-gu,
Seoul 120-749, Korea}

\KeyWords{Galaxy: center, ISM: molecules, ISM: supernova remnants}
\maketitle

\begin{abstract}
We present the statistical properties of molecular clumps in the Galactic center 50 km s$^{-1}$ molecular cloud (GCM-0.02-0.07) based on observations of the CS $J=1-0$ emission line with the Nobeyama Millimeter Array. In the cloud, 37 molecular clumps with local thermal equilibrium (LTE) masses of $2\times10^2-6\times10^3 M_\odot$ were identified by using the {\it clumpfind} algorithm. The velocity widths of the molecular clumps are about five-fold those of Galactic disk molecular clouds with the same radius. The virial-theorem masses are three-fold the LTE masses. The mass and size spectra can be described by power laws of $dN/dM\propto M^{-2.6\pm0.1}$ ($M\gtrsim 900M_\odot$) and $dN/dR\propto R^{-5.9\pm0.3}$ ($R\gtrsim 0.35$ pc), respectively. 
The statistical properties of the region interacting with the Sgr A East shell and those of the non-interacting part of the cloud are significantly different. The interaction probably makes the mass function steeper, from $dN/dM\propto M^{-2.0\pm0.1}$ in the non-interacting part to $dN/dM\propto M^{-4.0\pm0.2}$ in the interacting region. On the other hand, the interaction presumably truncates the size spectrum on the larger side of $R\sim 0.4$ pc.
\end{abstract}

\section{Introduction}
The Galactic center is the nearest nucleus of a spiral galaxy. The Central Molecular Zone (CMZ) (\cite{MorrisSerabyn}) is a molecular cloud complex extending along the Galactic plane up to $l\sim\pm1^\circ$ around the Galactic center. This complex is the counterpart of the central molecular cloud condensation often observed in nearby spiral galaxies. The molecular clouds in the CMZ are much denser, warmer, and more turbulent than disk molecular clouds (e.g. \cite {Bally1987}, \cite{Oka1998}, \cite{Tsuboi1999}). There are young and highly luminous star clusters in the CMZ,  such as Arches cluster, Quintuplet cluster, and the central cluster (e.g. \cite{Figer1999}, \cite{Figer2002}). The dense molecular clouds in the CMZ are the cradles of these bright star clusters. It is, however, an open question as to what mechanism is responsible for the formation of such star clusters in the CMZ. The 50 km s$^{-1}$ molecular cloud (GCM-0.02-0.07),  which is located only 3$'$ from Sagittarius A$^{\ast}$(Sgr A$^{\ast}$), is a most remarkable molecular cloud in the CMZ and harbors a massive active star-forming site (e.g. \cite{Ekers1983}, \cite{Goss1985}, \cite{Yusef-Zadeh2010}). The cloud may be forming such a bright star cluster.

As described in our previous paper (\cite{Tsuboi et al. 2009}, hereafter Paper I), we have observed the detailed structure of the 50 km s$^{-1}$ molecular cloud in the CS $J=1-0$ line with the Nobeyama Millimeter Array (NMA) of the Nobeyama Radio Observatory\footnote{Nobeyama Radio Observatory is a branch of the National Astronomical Observatory, National Institutes of Natural Sciences, Japan}. CS emission is expected to be almost free of strong contamination near 0 km s$^{-1}$ from the foreground and background disk molecular clouds because it has a high critical density, $n(H_2) \simeq 10^4$ cm$^{-3}$. The observation clearly shows the physical interaction between the 50 km s$^{-1}$ molecular cloud and the Sgr A East shell, which is a luminous and young supernova remnant (SNR) located in the region. This is consistent with observations of the OH 1720 MHz maser line (e.g. \cite{Sjouwerman2008}) and H$_2$ (1-0) S(1) emission line (e.g. \cite{Yusef-Zadeh2001}). The observation also shows that a young SNR is presumably embedded in the cloud, and the molecular gas shocked by the SNR is seen as a circle-like feature. The feature has a high temperature ratio of SiO and H$^{13}$CO$^+$ emission lines, $T($SiO$)/T($H$^{13}$CO$^+$) (see Figure 13 in \cite{Tsuboi2011}), because the SiO emission line is enhanced by shock. Similar molecular clouds interacting with SNRs must abound in the Galactic center region. The C-shock caused by the interaction might play an important role in star formation in the molecular clouds (e.g. \cite{Stahler}). The statistical properties of such shocked molecular clouds in the Galactic center region might be much different from those of disk clouds. Therefore, investigating these statistical properties should provide unprecedented information about the mechanism of star formation in the Galactic center region. 

Throughout this paper, we adopt 8.5 kpc as the distance to the Galactic center. At this distance, 1 pc corresponds to about $24\arcsec$. In addition, we use Galactic coordinates.

\begin{figure}[thp]
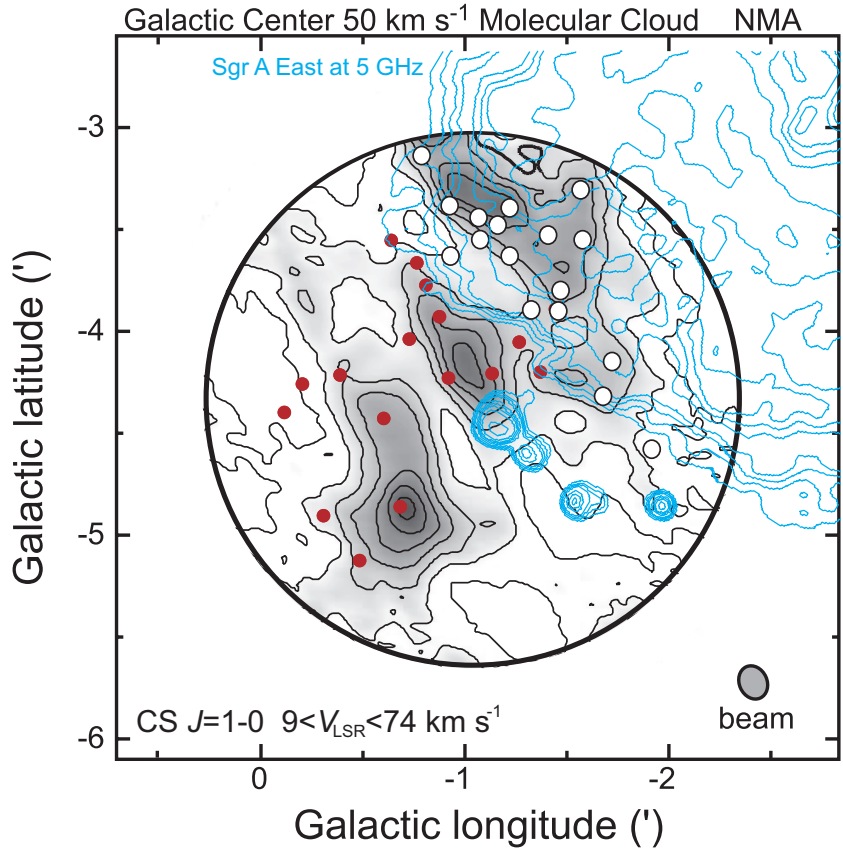

\begin{center}
\FigureFile(120mm,120mm){fig1.eps}
\caption{Velocity-integrated CS $J=1-0$ map of the 50 km s$^{-1}$ molecular cloud (paper I). The integrated velocity range is $V_{\rm{LSR}} = 9-74$ km s$^{-1}$. The size of the synthesized beam is $8.5\arcsec\times10\arcsec (\phi=24^\circ)$ and is shown in the bottom right corner. The thick circle shows the outer boundary of the FWHM, $D=156\arcsec$, of the element antenna of NMA. This figure is corrected for the primary beam attenuation of the element antenna. The lowest contour level and the contour interval are both $82~\mathrm{mJy~beam}^{-1}$.
Filled and open circles show the positions of the molecular clumps identified by the {\it clumpfind} algorithm (\cite{Williams1994}).  An image of Sgr A East in a 5 GHz continuum map (blue contours; \cite{Yusef-Zadeh1987}) is overlaid on the map. Open circles are located in the region interacting with Sgr A East (see Figure 3 in \cite{Tsuboi et al. 2009}).}
\label{Fig1}
\end{center}
\end{figure}

\begin{longtable}{lrrrrrrrrrrr}
\caption{Identified Molecular Clumps in the Galactic Center 50 km s$^{-1}$ Molecular Cloud }
\label{tbl:statistical}       
\hline
\hline
No.&$\Delta l$&$\Delta b$&$V_{\rm{LSR}}$&$T_{\rm{B}}$&$R$&$\Delta V$&$M_{\rm{LTE}}$& $M_{\rm{vir}}$&$\Delta V_{\rm{c}}$&$M_{\rm{LTE, c}}$& $M_{\rm{vir, c}}$\\
&[\arcsec]&[\arcsec]&[km s$^{-1}$]&[K]&[pc]&[km s$^{-1}$]&$[M_\odot]$&$[M_\odot]$&[km s$^{-1}$]&$[M_\odot]$&$[M_\odot]$\\
\hline
\endhead
\hline
\endfoot
\hline
\endlastfoot
1&-57.3& -3.4&75.6&5.7&0.18&5.9&1.6e+02&1.3e+03&3.9&5.0e+02&5.8e+02\\
2&-52.1& 5.1&71.8&5.1&0.20&10.1&1.0e+02&4.4e+03&6.8&3.2e+02&2.0e+03\\
3&-46.3 &29.7&56.5&5.0&0.38&11.8&2.1e+02&1.1e+04&7.9&6.8e+02&5.0e+03\\
4&-45.7 &-34.1 &29.7&5.1&0.25&9.5&2.0e+02&4.7e+03&6.4&6.2e+02&2.1e+03\\
5&-44.4& -9.0&75.6&4.2&0.25&4.3&1.0e+02&9.8e+02&2.9&3.3e+02&4.4e+02\\
6&-41.1&7.7&45.0&5.9&0.38&12.5&3.9e+02&1.2e+04&8.4&1.2e+03&5.6e+03\\
7&-35.2& -47.6&14.4&5.4&0.43&12.6&3.8e+02&1.4e+04&8.5&1.2e+03&6.5e+03\\
8&-27.9& -5.1&45.0&8.4&0.55&15.5&1.4e+03&2.8e+04&10.4&4.5e+03&1.3e+04\\
9&-25.8&47.7&60.3&5.9&0.50&7.6&5.7e+02&6.1e+03&5.1&1.8e+03&2.8e+03\\
10&-23.0 &-31.6 &29.7&10.4&0.66&13.5&2.0e+03&2.5e+04&9.1&6.3e+03&1.1e+04\\
11&-20.9&82.2&64.1&9.8&0.19&6.8&1.9e+02&1.8e+03&4.5&6.1e+02&8.2e+02\\
12&-20.3&18.4&60.3&5.1&0.45&10.0&4.8e+02&9.5e+03&6.7&1.5e+03&4.3e+03\\
13&-18.4&40.7&64.1&6.1&0.29&8.6&3.0e+02&4.5e+03&5.8&9.5e+02&2.0e+03\\
14&-17.3&72.9&67.9&8.0&0.32&9.3&3.5e+02&5.9e+03&6.2&1.1e+03&2.6e+03\\
15&-15.3&34.2&56.5&5.2&0.33&7.4&3.1e+02&3.8e+03&5.0&9.7e+02&1.7e+03\\
16&-11.7&24.8&37.3&6.6&0.38&19.2&6.2e+02&2.9e+04&12.9&1.9e+03&1.3e+04\\
17&-11.0& 83.2&56.5&8.9&0.20&10.0&1.4e+02&4.3e+03&6.7&4.5e+02&1.9e+03\\
18&-8.8&6.6&45.0&9.0&0.53&15.6&1.4e+03&2.7e+04&10.5&4.3e+03&1.2e+04\\
19&-7.8&57.7&60.3&7.7&0.31&10.5&3.6e+02&7.1e+03&7.1&1.2e+03&3.2e+03\\
20&-7.3&43.3&56.5&6.0&0.26&5.8&2.0e+02&1.8e+03&3.9&6.5e+02&8.2e+02\\
21&0.0&55.2&60.3&7.8&0.17&9.8&2.3e+02&3.5e+03&6.6&7.2e+02&1.6e+03\\
22&0.3&48.0&52.6&5.6&0.17&10.5&1.3e+02&3.9e+03&7.0&4.1e+02&1.8e+03\\
23&3.9&8.2&45.0&7.0&0.47&16.7&1.2e+03&2.8e+04&11.2&3.7e+03&1.3e+04\\
24&5.1&52.1&64.1&7.0&0.16&7.7&1.7e+02&2.0e+03&5.1&5.4e+02&9.1e+02\\
25&8.3&57.2&41.2&7.0&0.39&12.7&6.1e+02&1.3e+04&8.6&1.9e+03&6.0e+03\\
26&8.8&42.8&60.3&5.9&0.34&10.5&3.4e+02&7.8e+03&7.0&1.1e+03&3.5e+03\\
27&11.9&17.3&37.3&4.4&0.32&14.1&2.9e+02&1.4e+04&9.5&9.3e+02&6.1e+03\\
28&15.4&26.9&52.6&5.3&0.30&9.9&2.3e+02&6.1e+03&6.6&7.4e+02&2.8e+03\\
29&18.3&8.7&41.2&5.5&0.48&9.8&6.8e+02&9.6e+03&6.6&2.1e+03&4.3e+03\\
30&20.2&49.8&41.2&6.7&0.29&14.0&4.6e+02&1.2e+04&9.4&1.5e+03&5.4e+03\\
31&22.6&27.2&48.8&5.8&0.37&11.0&4.0e+02&9.5e+03&7.4&1.3e+03&4.3e+03\\
32&24.1&33.3&37.3&5.4&0.31&16.3&3.4e+02&1.7e+04&10.9&1.1e+03&7.7e+03\\
33&30.3&62.4&48.8&8.4&0.41&7.9&4.3e+02&5.4e+03&5.3&1.4e+03&2.4e+03\\
34&30.8&48.0&41.2&7.3&0.40&12.2&4.7e+02&1.2e+04&8.2&1.5e+03&5.6e+03\\
35&37.4&1.6&48.8&4.9&0.38&10.3&3.2e+02&8.5e+03&6.9&1.0e+03&3.8e+03\\
36&39.3&12.2&37.3&7.7&0.39&10.7&5.7e+02&9.4e+03&7.2&1.8e+03&4.2e+03\\
37&51.3&-14.0& 48.8&5.2&0.19&3.4&6.9e+01&4.5e+02&2.3&2.2e+02&2.0e+02\\%
\end{longtable}
%

\section{Data Analysis}
We have observed the 50 km s$^{-1}$ molecular cloud in CS $J=1-0$ emission line by using NMA (Paper I). The area mapped with NMA in the C$^{32}$S $J=1-0$ line (48.990964 GHz) is centered at $l=359^\circ58'54.8\arcsec, b=-0^\circ4'18.8\arcsec$. The full width at half-maximum (FWHM) of the element antenna of NMA is $156\arcsec$ at 49 GHz. The mapping area included the 50 km s$^{-1}$ molecular cloud. We obtained NMA data with 50 baselines. The velocity resolution of the channel map was 3.8 km s$^{-1}$. To make maps, we used the CLEAN method in the NRAO AIPS package. The size of the synthesized beam was $8.5\arcsec\times10\arcsec (\phi=24^\circ)$ for a natural weighting, which corresponds to about $0.35$ pc $\times 0.42$ pc at the Galactic center. Features with spatial scales larger than $1'$ or 2.5 pc were resolved out. A detailed description of the observation including the channel maps is given in Paper I. 
Figure 1 shows a velocity-integrated map of the 50 km s$^{-1}$ molecular cloud. The integrated velocity range is $V_{\rm{LSR}} = 9-74$ km s$^{-1}$, and covers almost all of the 50 km s$^{-1}$ molecular cloud. This figure is corrected for primary beam attenuation of the element antenna. The 50 km s$^{-1}$ molecular cloud was detected in the velocity range of $V_{\rm{LSR}} = 10.6-83.2$ km s$^{-1}$ (see also, Figure 2 in Paper I). The molecular cloud is resolved into many clumps and several filaments in the channel maps. 

 We used the {\it clumpfind} algorithm (\cite{Williams1994}) to automatically find candidates of the molecular clumps in the 50 km s$^{-1}$ molecular cloud. 
The {\it clumpfind} algorithm searches for emission peaks that have closed brightness surfaces (contours), and  that are isolated from the adjacent features, in the three-dimensional ($\alpha-\delta-v$) FITS data cube without making an assumption about the shape. 
The emission-free noise level of the data is $1\sigma = 0.462$~K in $T_B$i$=77$ mJy beam$^{-1}$).
In the scanning procedure of {\it clumpfind}, the contour increment and the lowest 
contour level are defined as adjustable parameters. We adopted 0.92~K ($\sim2\sigma$) 
as the contour increment and 2.31~K ($\sim5\sigma$) as the lowest contour level. 
The {\it clumpfind} algorithm found 64 clump candidates in the data cube. 
However, they still contain "fake" clumps (Cf. \cite{Pineda}).
Some candidates have an angular extent less than the size of the synthesized beam, or touch the boundary of the map. In addition, a few irregular candidates have low intensity 
and intricate extended structures.  To reject such candidates as having less validity,
we performed an inspection based on the same criterion used in the previous paper (\cite{MiyazakiTsuboi}).
 As the result, we identified 37 clumps in the 50 km s$^{-1}$ molecular cloud. 
The positions of the identified molecular clumps are shown as circles in Figure 1. 

The brightness temperatures of the identified molecular clumps are corrected for the primary beam attenuation of the element antenna. The brightness temperature ranges from 4.3 to 10.4 K with a mean of 6.5 K. To estimate the intrinsic radius of each clump, $R$, we subtracted the synthesized beam from the observed radius, $R_o$: 
$R=(R_{\rm o}^2-r_{\rm{beam}}^2)^{1/2}$, where $r_{\rm{beam}}$ is the average radius of the synthesized beam.
The estimated intrinsic radii range from 0.16 to 0.66 pc. Even the largest clump is not resolved out under the criterion mentioned above.
The positions, radial velocities, peak brightness temperatures,   estimated intrinsic radii, and FWHM velocity widths are summarized in Table 1.

The mass of molecular gas in the molecular cloud is estimated from the emission line intensity under the local thermal equilibrium (LTE) condition and an optically thin limit. This is the LTE molecular mass. 
The molecular column density based on CS $J=1-0$ observations is given by
\begin{equation}
\label{ }
N_{\mathrm{mol}}[\mathrm{cm}^{-2}]=\frac{7.55 \times 10^{11}T_{\mathrm{ex}}\int T_{\mathrm{B}}dv[\mathrm{K~km s}^{-1}]} {X({\mathrm{CS}})}.
\end{equation}
Here, $X({\mathrm{CS}})$ is the fractional abundance, $X({\mathrm{CS}})=N({\mathrm{CS}})/N_{mol}$, which is the relative abundance of CS molecules to total molecules; $T_{\mathrm{ex}}$ is the excitation temperature of CS molecules.
Previous NH$_3$ observations indicate that the rotational temperature of NH$_3$ in the 50 km s$^{-1}$ molecular cloud is $T_{\mathrm{rot.}} = 80$ K (\cite{Miyazaki2010}). The excitation temperature of relatively abundant molecule like CS has conventionally been considered to be $T_{\mathrm{ex}} = 50-60$ K in the Galactic center region because the excitation temperature is somewhat lower than the rotational temperature derived from NH$_3$ observation (e.g. \cite{Tsuboi1999}). Therefore, the excitation temperature of CS molecule is assumed to be $T_{\mathrm{ex}} = 50$ K. 

In the CMZ, $X({\mathrm{CS}})$ varies from cloud to cloud. The brightness temperature of the CS $J=1-0$ emission line and the temperature ratio of the CS $J=1-0$ and $J=2-1$ lines were observed as $T_{\mathrm{B}}({\mathrm{CS~1-0}})\sim 6$ K and $T_{\mathrm{B}}({\mathrm{CS~2-1}})/T_{\mathrm{B}}({\mathrm{CS~1-0}})\sim 0.9$ at the center of the 50 km s$^{-1}$ molecular cloud (\cite{Bally1987}, \cite{Tsuboi1999}).
According to the large velocity gradient model, the fractional abundance per velocity gradient of CS molecules is calculated as $X({\mathrm{CS}})/dv/dr\simeq 9\times10^{-10}$ pc/km s$^{-1}$ from the observed data. The velocity gradient is approximately the observed velocity width divided by the beam size, $dv/dr=8$ km s$^{-1}/0.8$ pc =10 km s$^{-1}/$pc. Thus, the estimated fractional abundance of CS molecules is $X({\mathrm{CS}})\simeq 1\times10^{-8}$ in the molecular cloud. The fractional abundance somewhat depends on the excitation temperature of CS molecules, but remains within several tens of percent of the estimated value in the range of $T_{\mathrm{ex}} = 30-70$ K. 

We calculated the LTE molecular mass of the cloud from the derived distribution of the molecular column density. The molecular mass is given by
\begin{equation}
\label{ }
M_{\mathrm{LTE}}[M_{\odot}] =\Omega[\mathrm{cm}^2]\mu[M_{\odot}] \sum_m\sum_n 
 N_{\mathrm{mol}}(m,n)[\mathrm{cm}^{-2}],
\end{equation}
where $\Omega$ is the physical area corresponding to the data grid, $\Omega=6.44 \times 10^{34} \mathrm{cm}^2 $ for a $2\arcsec$ grid spacing of the data cube and $\mu$ is the mean mass per one molecule with mean molecular weight of 2.3; thus, $\mu=1.94\times 10^{-57} M_{\odot}$. 

The mass of the molecular gas is also estimated from the size and velocity width assuming virial equilibrium. This is the virial-theorem mass of the molecular cloud, and is a good estimate at least in the disk region of the Galaxy.  In the case of an optically thin
emission line with a Gaussian profile, no external pressure, and no magnetic field, the virial-theorem mass of a spherical cloud with uniform density is nominally calculated as
\begin{equation}
\label{ }
M_{vir}[M_{\odot}]=210\times \Delta V[\mathrm{km~s}^{-1}]^2 R[\mathrm{pc}]
\end{equation}
where $ \Delta V$ is the FWHM of an emission line profile (e.g. \cite{MiyazakiTsuboi}). 
These masses of the clumps in the 50 km s$^{-1}$ molecular cloud are also given in Table 1.

Although this procedure is valid at the optically thin limit, a comparison of the C$^{32}$S $J=1-0$ and C$^{34}$S $J=1-0$ lines observed by the 45 m telescope of the Nobeyama Radio Observatory (NRO45) indicates that the averaged optical depth of the C$^{32}$S $J=1-0$ line reaches $\tau\simeq 3$ in the 50 km s$^{-1}$ molecular cloud (\cite{Tsuboi1999}). Line broadening will be caused by
such an optical depth. According to Phillips et al. (1979), the line broadening can be corrected
by
\begin{equation}
\label{ }
\Delta V_{\rm c}=ln2^{1/2}\Bigg\{ln\bigg\{\frac{\tau}{ln[2/(1+e^{-\tau})]}\bigg\}\Bigg\}^{-1/2}\Delta V,
\end{equation}
where $\Delta V_{\rm c}$ is the corrected FWHM velocity width.
When the optical depth  of the line is $\tau\simeq 3$, $\Delta V_{\rm c}= \Delta V/1.49$. 
The range of velocity width is from 2.3 to 12.9 km s$^{-1}$. The velocity widths are concentrated around the mean of 7.1 km s$^{-1}$. The corrected virial-theorem masses are calculated from these corrected velocity widths. 
On the other hand, the averaged optical depth gives the correction factor for the line intensity of $\tau/(1-e^{-\tau})=3.16$ in the 50 km s$^{-1}$ molecular cloud.
The LTE molecular masses at an optically thin limit are underestimated by this factor. 
The relation between $T_{\mathrm{B}}$ and $T_{\mathrm{ex}} $ is given by $T_{\mathrm{B}}\sim \eta T_{\mathrm{ex}}(1-e^{-\tau})$. The beam filling factor of the beam of $0.35$ pc $\times 0.42$ pc is estimated to be $\eta \sim 0.1$, assuming that $\tau\simeq 3$ and $T_{\mathrm{ex}}=50 $K. 
The corrected FWHM velocity widths, corrected LTE molecular masses, and corrected virial-theorem masses are also given in Table 1. 
In the following analysis, we use these corrected values.

\begin{figure}
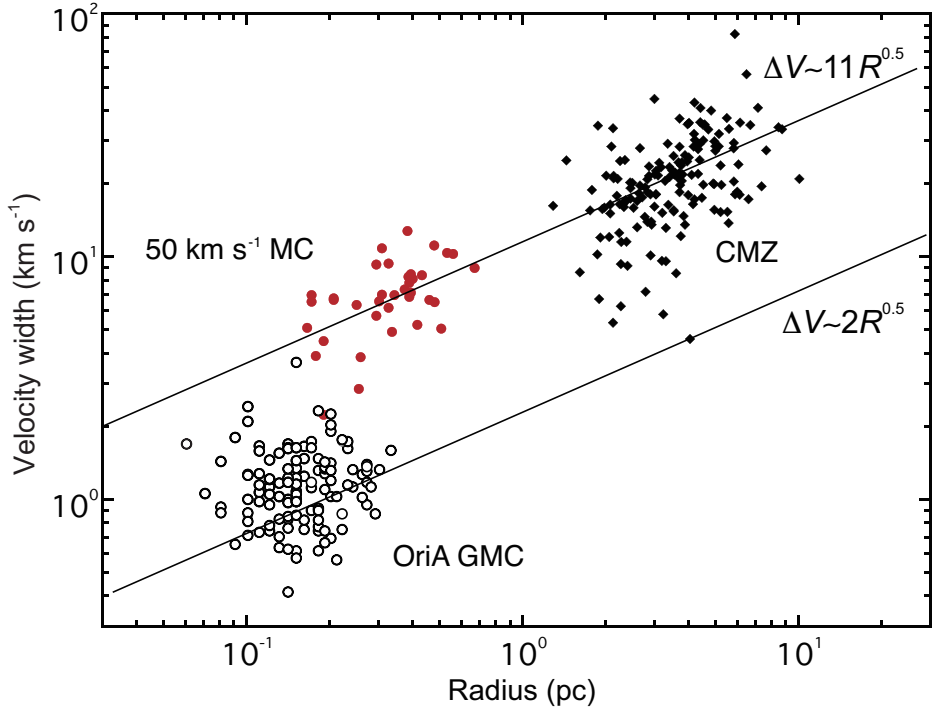

\begin{center}
\FigureFile(130mm,100mm){fig2.eps}
\caption{Velocity width--size relations of clumps in the 50 km s$^{-1}$ molecular cloud (filled circles), CMZ (diamonds \cite{MiyazakiTsuboi}), and OriA GMC (open circles \cite{Tatematsu}). The relations are from observations of the CS $J=1-0$ emission line. The sampling size in the 50 km s$^{-1}$ molecular cloud is a range from 0.16 to 0.66 pc. The velocity width--size relation of the clumps is bimodal. }
\label{Fig2}
\end{center}
\end{figure}

\section{Statistical Properties of the Identified Molecular Clumps}
\subsection{Velocity Width--Size Relation}
Figure 2 shows the velocity width--size relations of the molecular clumps in the 50 km s$^{-1}$ molecular cloud. 
The figure also shows those of the molecular clumps in the Orion A giant molecular cloud (hereafter OriA GMC) (\cite{Tatematsu}) and the CMZ (\cite{MiyazakiTsuboi}) for comparison. Both were observed in the CS $J=1-0$ emission line with NRO45. The observations of the same transition in the same molecular species are free from ambiguity derived from comparing different emission lines. The sampling size of this observation is
a range from 0.16 to 0.66 pc. The average size is $1/10$ that observed in the CMZ.

The velocity width--size relation is generally represented by $\Delta V \simeq aR^{-0.5}$ (\cite{Larson}). 
The velocity width--size relations of the disk molecular clouds, including that of Ori A GMC, are usually taken as $\Delta V \simeq 2R^{-0.5}$ (see also, \cite{Solomon}). Therefore, there is a common velocity width--size relation in the disk region of the Galaxy. On the other hand, the velocity width--size relation in the CMZ is $\Delta V \simeq 11R^{-0.5}$ (\cite{MiyazakiTsuboi}). 
However, it is not clear whether the velocity width--size relation is bimodal or the relation depends on  clump size, because the clump sizes identified in the CMZ are over one order of magnitude larger than those in the OriA GMC. In the present observations, the average size of the identified clumps is only two-fold that in the OriA GMC and the size ranges of the two cases overlap to a fair extent. 
The identified clumps in the 50 km s$^{-1}$ molecular cloud follow the same relation, but at smaller size, as clumps in the CMZ. The velocity widths of the CMZ and the 50 km s$^{-1}$ molecular cloud are about five-fold those of disk molecular clouds having the same radius. There is presumably another common velocity width--size relation, $\Delta V \simeq 11R^{-0.5}$, in the CMZ and the 50 km s$^{-1}$ molecular cloud, although the clumps in the 50 km s$^{-1}$ molecular cloud would be influenced by the incident shock wave. 

\begin{figure}
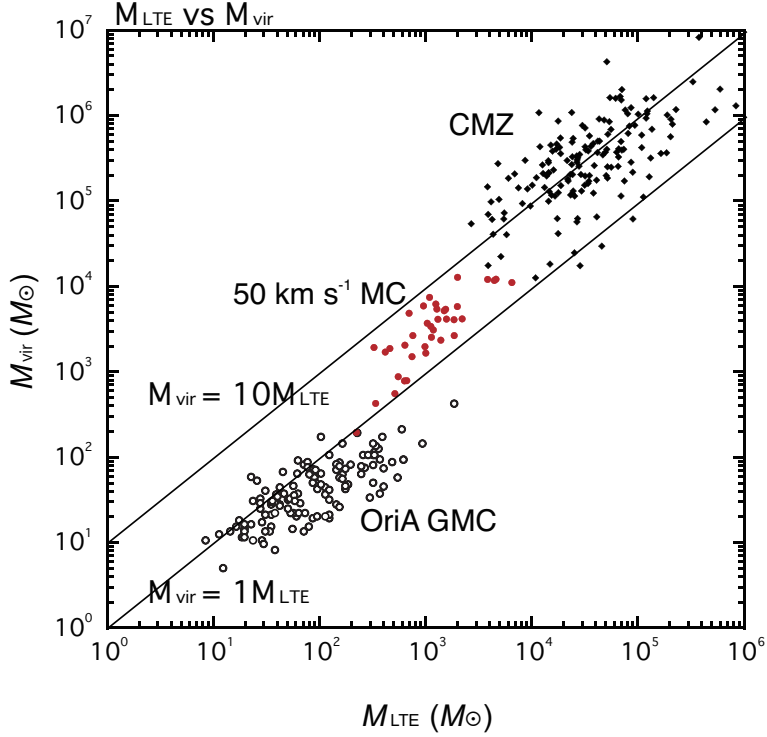

\begin{center}
\FigureFile(110mm,50mm){fig3.eps}
\caption{Relations between the LTE mass and virial-theorem mass of clumps in the 50 km s$^{-1}$ molecular cloud (filled circles), CMZ (diamonds \cite{MiyazakiTsuboi}), and OriA GMC (open circles \cite{Tatematsu}).} 
\label{Fig4}
\end{center}
\end{figure}

\subsection{Relation between LTE mass and virial-theorem mass}
Figure 3 shows the relation between the LTE mass and virial-theorem mass of the molecular clumps in the 50 km s$^{-1}$ molecular cloud. The relations for the OriA GMC and the CMZ are also shown in the figure for comparison. The relations for the OriA GMC indicates that the virial-theorem masses are nearly equal to the LTE masses (see also, Figure 10 in \cite{Tatematsu}). These results show that the molecular cloud clumps in the GMC are near virial equilibrium. On the other hand, the virial-theorem masses in the CMZ are an order of magnitude larger than the LTE masses (\cite{MiyazakiTsuboi}). The discrepancy between the virial-theorem masses and the LTE masses suggests that the clumps identified in the CMZ are far from virial equilibrium because the discrepancy is larger than the uncertainty in the LTE mass from the fractional abundance. Oka et al. have suggested that external pressure is required for binding the clumps (\cite{Oka2001}). The poloidal magnetic field of the Galactic center region is believed to be widely uniform and as strong as 1 mG (\cite{Yusef-Zadeh1987}, \cite{Yusef-Zadeh1987b}). Such a magnetic field may be a source of the external pressure. 

However, the clumps identified in the CMZ are three orders of magnitude more massive than those in the OriA GMC. This large difference in mass obscures whether or not the relation is bimodal. The clumps identified in the 50 km s$^{-1}$ molecular cloud are only one order of magnitude greater in LTE mass than those in the OriA GMC. In the 50 km s$^{-1}$ molecular cloud, the virial-theorem masses of the clumps are three-fold the LTE masses, $M_{vir}\simeq 3\times M_{\rm LTE}$. The proportional constant in the 50 km s$^{-1}$ molecular cloud is just between those of the two previous cases. 

Any external pressure or other binding mechanism is still required if the discrepancy is real in this case. The magnetic field is well-ordered in the 50 km s$^{-1}$ molecular cloud (\cite{Novak2000}), and is aligned with the elongations in the cloud structures (see Figure 7 in Paper I). This field is not aligned along the large-scale poloidal magnetic field mentioned above, and is not parallel to the Galactic plane, although the large-scale magnetic field in the CMZ is oriented approximately parallel to the Galactic plane (\cite{Chuss2003}). 
Therefore, the specific magnetic field presumably surrounds the cloud. The magnetic field may bind the partially ionized molecular gas in the cloud. 

An alternative explanation of the discrepancy between the LTE mass and the virial-theorem mass is overestimation of the virial-theorem mass. This could originate from overestimation of the effective radius of the clumps (see Eq. 3). The observed radius of the clumps, which should be similar to the radius of the surface of $\tau=2/3$ (photosphere) in the CS $J=1-0$ emission line, is presumably larger than the effective radius because the clumps are optically thick, $\tau\simeq 3$ (see Sec. 2). In addition, a small beam-filling factor may also cause the overestimation. If the beam-filling factor is as small as $0.1$, as mentioned previously, the virial-theorem mass  decreases to $\sqrt{0.1}=30$ \%. In such a case, the external pressure is no longer required to reach equilibrium.
In either case, this observations is not conclusive for the binding mechanism of the clumps because of its low angular resolution. The Atacama Large Millimeter-submillimeter Array (ALMA)
should identify many clumps in the cloud and image their fine structures. This information will be crucial in resolving this issue. 

\begin{figure}
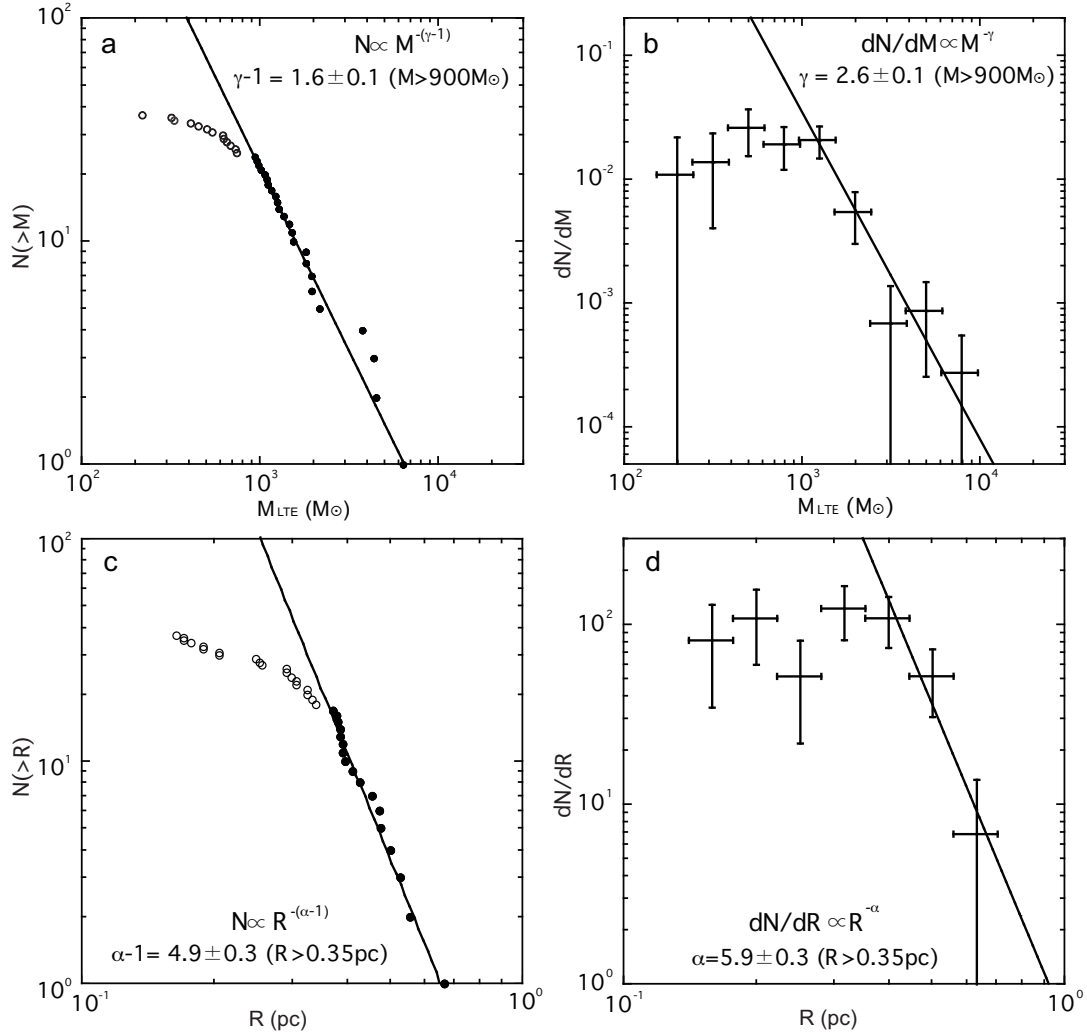

\begin{center}
\FigureFile(160mm,160mm){fig4.eps}
\caption{{\bf a} Cumulative mass function in the 50 km s$^{-1}$ molecular cloud, which is the relation between the LTE mass of the molecular clumps and the accumulated number of the clumps with mass greater than the LTE mass. The filled circles show data points with the turn-over mass of $M\simeq 900M_\odot$. The inclined line shows the best-fit power law, $N\propto M^{-1.6\pm0.1}$. {\bf b} Mass function of clumps in the 50 km s$^{-1}$ molecular cloud. The inclined line is derived from the line shown in {\bf a}, $dN/dM\propto M^{-2.6\pm0.1}$. 
 {\bf c} Cumulative size spectrum in the 50 km s$^{-1}$ molecular cloud, which is the relation between the estimated intrinsic radii of the molecular clumps and the accumulated number of the clumps with size greater than the intrinsic radius. The filled circles show data points with the turn-over radius of $R\simeq 0.35$ pc. The inclined line shows the best-fit power law, $N\propto R^{-4.9\pm0.3}$. {\bf d} Size spectrum of clumps in the 50 km s$^{-1}$ molecular cloud. The inclined line is derived from the line shown in {\bf c}, $dN/dR\propto R^{-5.9\pm0.3}$.}
\label{Fig5}
\end{center}
\end{figure}

\subsection{Mass function and size spectrum}
\subsubsection{Whole 50 km s$^{-1}$ molecular cloud}
The mass function of the clumps should provide important information about the molecular cloud, and can often be described by a power law:
\begin{equation}
\label{ }
dN/dM\propto M^{-\gamma},
\end{equation}
where $M$ is the mass of a clump and $N$ is the number of clumps with mass of $M\sim M+dM$. 
The cumulative mass function is calculated by integration of this relation:
\begin{equation}
\label{ }
N(>M)=\int^\infty_M(dN/dM)dM=aM^{-(\gamma-1)}+C.
\end{equation}
This equation describes the relation between the LTE mass of the molecular clumps and the total number of the clumps, $N$, with mass larger than $M$, that is, $N(>M)$. 
Here, we use the cumulative mass function to derive the effective mass function. Figure 4a shows the cumulative mass function of the clumps in the 50 km s$^{-1}$ molecular cloud. The sampling mass of this observation is in the range of $M_{\mathrm{LTE}}=2\times10^2-6\times10^3 M_\odot$. 
Although the relation is warped at higher mass and curved at lower mass, the main part above about 900 $M_\odot$ can be described by a power law, $N(>M)\propto M^{-(\gamma-1)}$. 
At higher mass, the relation may be warped by source confusion. At lower mass, the relation is affected by limited sensitivity. 
The power law index is derived as $\gamma-1 =1.6\pm0.1$ by curve-fitting. 
Figure 4b shows the mass function of clumps in the 50 km s$^{-1}$ molecular cloud. 
The power law index of the straight line in the figure is $\gamma=2.6\pm0.1$ ($M\gtrsim 900M_\odot$). 

Through the same procedure, the power law index of the mass function in the CMZ is re-calculated to be $\gamma=1.9\pm0.1$ ($M\gtrsim 2\times 10^4M_\odot$) from the CS survey (\cite{Tsuboi1999}), whereas our original value was $\gamma=1.6\pm0.1$ ($M\gtrsim 1\times 10^4M_\odot$)(\cite{MiyazakiTsuboi}). A recent sub-millimeter observation using Bolocam of the Caltech Submillimeter Observatory shows that the mass function for $M\gtrsim 80 M_\odot$ is a power law with an index of $\gamma\simeq 2.1$ in the Galactic center region including the CMZ (\cite{Bally}). The power law index derived in the 50 km s$^{-1}$ molecular cloud is somewhat larger than these indexes obtained for the molecular clouds in the CMZ. However, these previous observations have larger beam sizes ($60\arcsec$ for the CS survey and $33\arcsec$ for the Bolocam survey). Because the power law indexes of the mass function (and also size spectra) strongly suffer from source confusion, especially in a crowded region such as the CMZ, it is not yet conclusive whether the index in the 50 km s$^{-1}$ molecular cloud is different from those in the CMZ. 

We also calculate the size spectrum for the clumps by using the same procedure. 
 Figure 4c shows the cumulative size spectrum of the clumps in the 50 km s$^{-1}$ molecular cloud. Although the sampling size range of this observation is only $R=0.16-0.66$ pc, the cumulative size spectrum in the range over $R\gtrsim 0.35$pc can be expressed as a power law, $N\propto R^{-(\alpha-1)}$.  The power law index is derived as $\alpha-1 =4.9\pm0.3$ by curve-fitting. 
Figure 4d shows the size spectrum of clumps in the 50 km s$^{-1}$ molecular cloud. The inclined line in the figure is derived from the line shown in figure 4c, $dN/dR\propto R^{-5.9\pm0.3}$($R\gtrsim 0.35$ pc).
The power index of the CMZ is derived as $\alpha = 4.1\pm0.4$ ($R\gtrsim 3.5$ pc) by using the same procedure as in the CS survey (\cite{MiyazakiTsuboi}). The derived power law index in the 50 km s$^{-1}$ molecular cloud is notably larger than that in the CMZ. 
 
 \subsubsection{Difference between interacting and non-interacting regions}
 The observed area contains the molecular cloud interacting with Sgr A East. The statistical properties of such a region might be much different from those of disk clouds.
Next, we will search for differences in the statistical properties of the region interacting with Sgr A East and the non-interacting part in the cloud. 
As shown in Figure 1, 19 clumps are identified in the interacting region with Sgr A East, and these clumps are in the mass range of $< 2000M_\odot$. Figure 5a shows the cumulative mass function in the region interacting with Sgr A East. The mass function above about 900 $M_\odot$ can be described by a power law. The inclined line shows the best-fit power law, $N\propto M^{-3.0\pm0.2}$.
 Then, the power law index of the mass function is derived as $\gamma=4.0\pm0.2$. The power law index in the interacting region is notably steeper than the indexes obtained for the above-mentioned molecular clouds with larger mass and larger size in the CMZ and in the disk region of our Galaxy, for example, $\gamma=2.3\pm0.2$ in Orion B (\cite{Ikeda2009a}), $\gamma=2.3\pm0.3$ in OMC-1 (\cite{Ikeda2009b}), and $\gamma=2.5\pm0.2$ in the Gould Belt(\cite{Andre}). The interaction of the molecular clouds with SNRs probably makes the mass function steeper. 
\begin{figure}
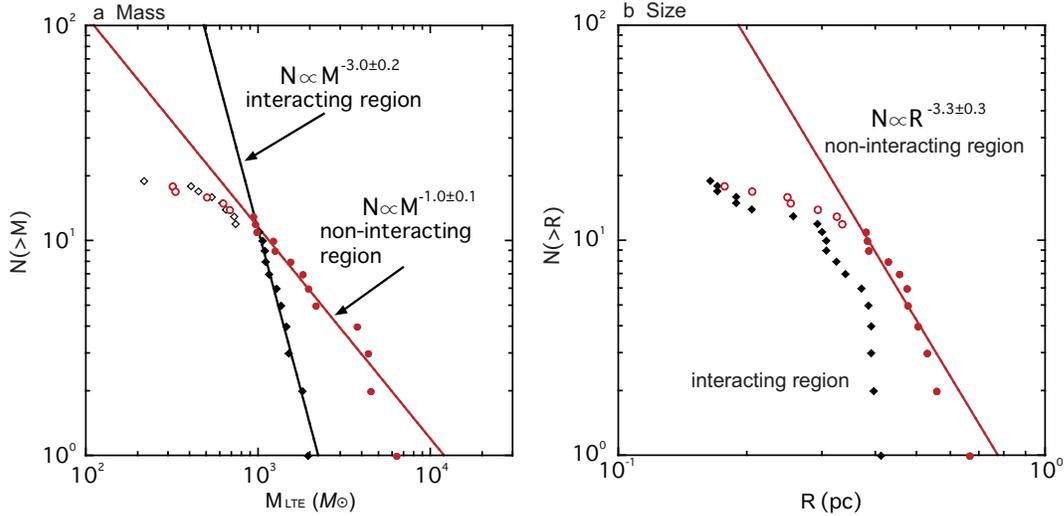

\begin{center}
\FigureFile(160mm,70mm){fig5.eps}
\caption{{\bf a} Cumulative mass functions of the molecular clumps in the region interacting with Sgr A East (filled and open diamonds) and non-interacting regions (filled and open circles) (see also, Figure 1). The inclined lines show best-fit power laws, $N\propto M^{-3.0\pm0.2}$ for the interacting region and $N\propto M^{-1.0\pm0.1}$ for the non-interacting regions. {\bf b} 
Cumulative size spectra in the interacting region (filled diamonds) and non-interacting regions (open and filled circles). The size spectrum of the interacting region is truncated on the right of $R\sim 0.4$ pc. On the other hand, the power law index of the non-interacting region is derived as $\alpha = 4.3\pm0.3$ ($R\gtrsim  0.35$ pc), which is approximately equal to the power law index of the CMZ. 
}
\label{Fig5}
\end{center}
\end{figure}

Figure 5a also shows the cumulative mass function of non-interacting molecular clumps in the cloud. In the region, 18 clumps are identified. The cumulative mass functions of the interacting and non-interacting regions are remarkably different. The cumulative mass function above about 900 $M_\odot$ can be also described by a power law, $N\propto M^{-1.0\pm0.1}$. 
The derived power law index of the mass function is $\gamma=2.0\pm0.1$, which is approximately equal to the indexes obtained in the CMZ and in the disk region of our Galaxy. As mentioned above, the power law indexes in the disk region are usually in the range of $\gamma\sim 2.0-2.5$. A similar index is also seen in the non-interacting molecular clumps in the Galactic center region. 

Figure 5b shows the cumulative size spectra of the clumps in the region interacting with Sgr A East and in the non-interacting part in the 50 km s$^{-1}$ molecular cloud. There is also a remarkable difference between the size spectra of the interacting and non-interacting regions. The size spectrum of the interacting region is presumably truncated on the right of $R\sim 0.4$ pc. The interaction of molecular clouds with SNRs probably changes the size spectrum. On the other hand, the power law index of the non-interacting part in the 50 km s$^{-1}$ molecular cloud is derived as $\alpha = 4.3\pm0.3$ ($R\gtrsim   0.35$ pc), which is approximately equal to the power law index of the CMZ mentioned above. 

 The mass function and size spectrum observed in the interacting region indicate that over 50\% of the clumps have similarly high mass, $M\sim 1400\pm 300 M_\odot$, and similarly small size, $R\sim 0.38\pm 0.06$ pc. The interacting region may be somewhat homogenized by its restricted physical properties compared with those of other regions. For example, the spatial scale containing the clumps may be restricted by the thickness of the shocked gas. The diversity of the mass and size of the clumps is probably lowered as a consequence. The clumps with restricted mass and size may play an important role in the formation of star clusters with many high mass stars. This suggests that the interaction produces a top-heavy initial mass function (IMF). The young and highly luminous clusters in the Galactic center environment are believed to have such IMF. However, the number of the identified clumps in this observation is not sufficient to derive an indisputable mass function of the interacting region. ALMA should resolve this issue by finding many more clumps in the cloud.
\section{Conclusions}
We have presented the statistical properties of the Galactic center 50 km s$^{-1}$ molecular cloud (GCM-0.02-0.07) based on observations of CS $J=1-0$ with NMA. In the cloud, 37 molecular clumps were identified by using the {\it clumpfind} algorithm. The velocity width--size relation of the clumps is bimodal. In the CMZ, this relation is $\Delta V \simeq 11R^{-0.5}$, whereas in Galactic disk molecular clouds, the relation is $\Delta V \simeq 2R^{-0.5}$. The virial-theorem masses are three-fold the LTE masses, $M_{vir}\simeq 3\times M_{\rm LTE}$.
The mass and size spectra can be described by power laws of $dN/dM\propto M^{-2.6\pm0.1}$ ($M\gtrsim 900M_\odot$) and $dN/dR\propto R^{-5.9\pm0.3}$ ($R\gtrsim 0.35$ pc), respectively. 
There are significant differences in the statistical properties of the region interacting with Sgr A East and the non-interacting part in the cloud. The interaction probably makes the mass function steeper, from $dN/dM\propto M^{-2.0\pm0.1}$ in the non-interacting part to $dN/dM\propto M^{-4.0\pm0.2}$ in the interacting region. On the other hand, the interaction presumably truncates the size spectrum on the larger side of $R\sim 0.4$ pc.

\bigskip 
The authors would like to thank Dr. N. Ikeda and Prof. Y. Kitamura at the Institute of Space and Astronautical Science for useful discussions. We also thank Prof. S. K. Okumura at National Astronomical Observatory of Japan for useful discussions in the initial phase of this study.

\end{document}